\documentclass[conference]{IEEEtran}
\IEEEoverridecommandlockouts

\usepackage{cite}
\usepackage{amsmath,amssymb,amsfonts}
\usepackage{graphicx}
\usepackage{xcolor}
\usepackage[colorinlistoftodos]{todonotes}
\usepackage{booktabs}
\usepackage{float}
\usepackage{placeins}
\usepackage{multirow}
\usepackage[normalem]{ulem}
\usepackage[framemethod=TikZ]{mdframed}
\usepackage{verbatim}
\usepackage{listings}
\usepackage{adjustbox}
\usepackage{tabularx}
\usepackage{url}
\usepackage{xurl}
\usepackage[breaklinks=true]{hyperref}
\usepackage{siunitx}
\definecolor{darkgreen}{rgb}{0.0, 0.5, 0.0}
\usepackage{caption}
\usepackage{subcaption}

\begin{document}

\title{%
  \vspace{-1.5cm} 
  {\normalfont\normalsize\centering
\small 3rd IEEE International Conference on AI in Cybersecurity (ICAIC), DOI: \href{https://doi.org/10.1109/ICAIC60265.2024.10433839}{10.1109/ICAIC60265.2024.10433839}, PREPRINT COPY\\
  \vspace{0.5cm} 
  }
  CANAL - Cyber Activity News Alerting Language Model : Empirical Approach vs. Expensive LLMs
}

\author{\IEEEauthorblockN{Urjitkumar Patel}
\IEEEauthorblockA{\textit{Ratings Data Science} \\
\textit{S\&P Global}\\
New York, USA \\
urjitkumar.patel@spglobal.com}
\and
\IEEEauthorblockN{Fang-Chun Yeh}
\IEEEauthorblockA{\textit{Ratings Data Science} \\
\textit{S\&P Global}\\
New York, USA \\
jessie.yeh@spglobal.com}
\and
\IEEEauthorblockN{Chinmay Gondhalekar}
\IEEEauthorblockA{\textit{Ratings Data Science} \\
\textit{S\&P Global}\\
New York, USA \\
chinmay.gondhalekar@spglobal.com}
}
\maketitle
\IEEEpubidadjcol
\begin{abstract}
In today's digital landscape, where cyber attacks have become the norm, the detection of cyber attacks and threats is critically imperative across diverse domains. Our research presents a new empirical framework for cyber threat modeling, adept at parsing and categorizing cyber-related information from news articles, enhancing real-time vigilance for market stakeholders. At the core of this framework is a fine-tuned BERT model, which we call CANAL - Cyber Activity News Alerting Language Model, tailored for cyber categorization using a novel silver labeling approach powered by Random Forest. We benchmark CANAL against larger, costlier LLMs, including GPT-4, LLaMA, and Zephyr, highlighting their zero to few-shot learning in cyber news classification. CANAL demonstrates superior performance by outperforming all other LLM counterparts in both accuracy and cost-effectiveness. Furthermore, we introduce the Cyber Signal Discovery module, a strategic component designed to efficiently detect emerging cyber signals from news articles. Collectively, CANAL and Cyber Signal Discovery module equip our framework to provide a robust and cost-effective solution for businesses that require agile responses to cyber intelligence.
\end{abstract}

\begin{IEEEkeywords}
Large Language Models (LLM), BERT, Natural Language Processing (NLP), Machine Learning, Generative AI (GenAI), Cyber Risk Modeling, Cyber Signal Discovery, Cyber News Alerts, Empirical Cost Analysis
\end{IEEEkeywords}
    \vspace{2mm}

\section{Introduction}

The recent escalation in cyber attacks is a significant concern that spans across multiple sectors. As reported by ISACA’s 2023 State of Cybersecurity \cite{isaca_state_of_cybersecurity_2023}, there's an evident increase in cyberattacks across organizations. This trend is not limited to one domain; it affects diverse areas such as technology, oil and gas, healthcare, education, and finance. The technology sector frequently confronts data breaches and intellectual property theft, while the oil and gas industry faces threats to its critical infrastructure. In finance, cyber attacks can lead to significant financial losses and undermine consumer confidence. Each of these sectors, including healthcare, where data encryption in ransomware attacks is alarmingly high \cite{SophosHealthcare2023}, and education, with its highest rate of ransomware incidents \cite{SophosEducation2023}, demonstrates the broad spectrum of cyber vulnerability.

In this context, the timely detection and dissemination of information about cyber attacks and threats become crucial. News articles, in particular, play an essential role in providing real-time information and early warnings. We define a framework which combines entity relevance with a novel categorization scheme that segments cyber-related news into five distinct groups, each providing unique insights. This system not only aids in constructing comprehensive cyber profiles for entities but also facilitates informed decision-making by providing nuanced insights into their cyber risk exposure.

Recent advancements in Natural Language Processing (NLP), especially transformer-based models like BERT \cite{devlin2019bert}, T5 \cite{raffel2023exploring}, Flan \cite{wei2022finetuned}, GPT \cite{openai2023gpt4}, Llama \cite{touvron2023llama}, Mistral \cite{jiang2023mistral}, Claude \cite{anthropic_model_card_2023}, PaLM \cite{chowdhery2022palm} have revolutionized various NLP tasks, including applications in the cyber domain. In past year, we have seen capable applications such as OpenAI's ChatGPT \cite{ChatGPT}, Google's Bard \cite{Bard} which are powered by multi billion parameter models, GPT4 and Palm respectively, have become integral part of indiviuals and organizations for their daily tasks. These powerful models have shown remarkable capabilities in processing and understanding complex language structures. However, the widespread adoption of these advanced models is often hampered by their need for substantial computational resources and the associated high costs of infrastructure or API usage. We demonstrate that for tasks such as Cyber Categorization, simpler and smaller BERT \cite{devlin2019bert} model when finetuned with good data, outperforms these huge expensive models.

Addressing the prevalent challenges, our research introduces a cost-efficient empirical framework, leveraging a finetuned BERT architecture with minimal training data requirements. This framework excels in accurately categorizing cyber-related content from news articles, ensuring timely awareness for stakeholders. A key feature is the Cyber Signal Discovery module, adept at spotting emerging threats and enhancing our cyber terminology database with human-verified updates. Below, we outline the pivotal contributions of our research to the field:

\vspace{5mm}
\begin{itemize}
    \item \textbf{Introduction of CANAL (Cyber Activity News Alerting Language Model):} A cutting-edge model specifically developed for the efficient categorization of cyber-related information within news articles.
    \vspace{5mm}
    \item \textbf{Unique 5 Class Cyber Categorization:} We introduce a novel classification scheme, grouping cyber-related news into five distinct categories, each serving a different function across business domains.
    \vspace{1.5mm}
    \item \textbf{Efficient Fine-Tuning of CANAL:} Empirical fine-tuning framework using a minimal dataset, achieving state-of-the-art results.
    \vspace{1.5mm}
    \item \textbf{Benchmarking Against Larger Models:} Comparative analysis with major LLMs, focusing on zero to few-shot learning in cyber classification.
    \vspace{1.5mm}
    \item \textbf{Cyber Signal Discovery Module:} Advanced module for detecting emerging cyber threats, integrated with human expertise.
    \vspace{1.5mm}
    \item \textbf{Cost-Effective and Resource-Efficient Solution:} Demonstrating a more economical and efficient alternative in cyber categorization.
    \vspace{1.5mm}
    \item \textbf{Cyber Alerting Solution that Utilizes News Data:} Utilizing raw news data for comprehensive and up-to-date cyber threat alerting.
    \vspace{2mm}
\end{itemize}

\section{Literature Review}

The realm of cybersecurity is undergoing a significant transformation, driven by the increasing frequency and complexity of cyber threats in various sectors. The ISACA 2023 State of Cybersecurity report indicates a rise in cyberattacks impacting sectors such as technology, healthcare, education, and finance \cite{isaca_state_of_cybersecurity_2023}. In response to these challenges, Artificial Intelligence (AI) is playing a crucial role in evolving the cybersecurity landscape. AI-driven approaches are enhancing threat detection capabilities and enabling automated responses.

The advent of transformer-based models, such as those introduced by Vaswani et al. \cite{vaswani2023attention}, has ushered in a new era in NLP. Jacob Devlin et al.'s BERT \cite{devlin2019bert} stands out in this evolution, leveraging deep bidirectional representations to transform tasks like question answering and language inference. Its application in domain-specific tasks, as exemplified by Dogu Araci's FinBERT \cite{araci2019finbert}, showcases the adaptability of encoder-only transformer models, which are particularly effective for cyber-specific classification tasks where generative capabilities are less critical.

While generative models like llama \cite{touvron2023llama}, BloombergGPT \cite{wu2023bloomberggpt}, GPT \cite{openai2023gpt4} \cite{brown2020language} have expanded the scope of transformers, their scalability, seen in models like PaLM \cite{chowdhery2022palm}, introduces significant resource demands. The resource-intensive nature of these models \cite{Ganesh_2021} and the associated costs \cite{floridi2020gpt3} make them less practical for smaller, specific applications. Hoffman et al.'s Chinchilla study \cite{hoffmann2022training} offers a balanced perspective on model size and training data, yet the deployment of such large models in constrained environments remains debatable. This context underscores the relevance of models like BERT, which provide a more feasible and efficient approach for specialized tasks in the realm of cybersecurity.

The integration of NLP into cybersecurity, particularly using transformer-based models like BERT, has opened new avenues in threat detection and classification. Notable applications of BERT in cybersecurity extend beyond traditional models. Ebelechukwu et al.'s CAN-BERT for anomaly detection in automotive systems \cite{10069190}, Rahali et al.'s MalBERT for Malware identification using Android application source code \cite{bdcc7020060}, and Chen et al.'s BERT-Log for system log interpretation \cite{doi:10.1080/08839514.2022.2145642} are a few examples. Adding to this, Kimia Aneri et al. introduced CyBERT, which fine-tunes BERT for cybersecurity claim classification, focusing on optimal hyperparameters for enhanced accuracy \cite{jcp1040031}. Furthermore, Salih Yasir et al. explored malware detection using fastText and BERT, emphasizing the purification and optimization of API call sequences for classification tasks \cite{9486377}. Moreover, Aghaei et al. introduced SecureBERT, a cybersecurity language model capable of capturing text connotations in cybersecurity text tailored for Cyber Threat Intelligence (CTI) tasks, exemplifying the model's adaptability to specialized cybersecurity needs \cite{aghaei2022securebert}. These diverse applications underscore BERT's adaptability and effectiveness in addressing specialized cybersecurity challenges.

The evolution of cybersecurity alert systems has been marked by significant advancements through the analysis of online data. In 2015, Yigit Erkal et al. utilized Twitter data for cyber security content identification, using a Naive Bayes Classifier and TF-IDF vectorization \cite{7424414}. By 2018, Mohamad Syahir Abdullah et al. furthered this field by focusing on detecting cyber-attack news, employing a Conditional Random Field classifier and Latent Semantic Analysis \cite{electronics11020198}. In 2021, Thea Riebe et al. introduced CySecAlert, a Twitter-based system for alerting on cyber security topics \cite{10.1007/978-3-030-86890-1_24}. On a cyber specific NER, Md TanvirAlum et al. proposed CyNER, a Python library for cybersecurity named entity recognition (NER), leveraging transformer-based models and heuristics to extract cybersecurity-related entities \cite{alam2022cyner}

However, despite these advancements, there appears to be a gap in the literature concerning the application of BERT-based models for generating cyber alerts from more real time news sources like Google News. To the best of our knowledge, no studies have yet explored the finetuning of BERT models specifically for the generation of cyber alerts based on Google News alert data. This gap signifies a potential area for future research, particularly given BERT's proven efficacy in text classification and its potential applicability in the nuanced field of cyber risk modeling.

Our study addresses a significant gap in cyber threat detection by introducing a framework that generates entity-focused cyber alerts from online news, utilizing cost-effective and efficient Large Language Models (LLMs). This approach overcomes key challenges such as limited training data and the dynamic nature of cyber threats. By processing vast amounts of online news to identify relevant cyber risks, our framework enhances the accuracy and timeliness of alerts, making it a valuable addition to existing cyber threat intelligence systems. Overall, our research contributes to advancing cyber threat detection strategies in today's digital landscape.

\section{Background And Theory}

\subsection{Problem Statement}

In the rapidly evolving domain of cybersecurity, the daily influx of thousands of news articles presents a significant challenge in information management and analysis. Our study's goal is two-fold: firstly, to effectively categorize these articles into five distinct categories — Recent Cyber Attack, Litigation, Future Threats, Preventive Action, and ``Other''; and secondly, to discover and highlight emerging cyber threats and signals. This dual approach is essential for systematically organizing and analyzing the vast array of incoming cyber-related news, enabling a more focused and efficient method for cyber threat intelligence. Our system aims to achieve these objectives using a practical, efficient, and cost-effective approach, distinguishing it from the more expensive Large Language Models (LLMs). The categorization is as follows:

\begin{itemize}

\item \textbf{Recent Cyber Attack :}
This category encompasses articles that report on recent real cyber attacks targeting entities. These articles provide critical insights into actual cyber threats that have resulted in tangible damage to companies. These Articles constitute a significant portion of our dataset, as they offer invaluable information for understanding the evolving cyber threat landscape.

\item \textbf{Cyber Litigation :}
Litigation Articles pertain to news articles that discuss legal actions, investigations, or charges related to cyber incidents. These articles are crucial for professionals tracking legal developments and consequences in the cybersecurity domain.

\item \textbf{Future Cyber Threats:}
Future Threats category includes articles that address potential cyber risks and threats that organizations may face in the future. These articles are forward-looking and target an audience interested in proactive risk assessment.

\item \textbf{Cyber Risk Preventive:}
Preventive Action articles highlight positive actions, remedies, vulnerability fixes, and patches aimed at reducing the likelihood of future cyber risks. This category is particularly important as it contributes to building a positive cyber profile for an entity.

\item \textbf{Other:}
The ``Other'' category encompasses a diverse range of articles, including reports, studies, educational content, guidance materials, and miscellaneous topics related to cybersecurity. This category serves as a catch-all for articles that do not neatly fit into the first four classifications, providing additional context and insights into various aspects of cybersecurity.

\end{itemize}

By efficiently sorting these articles into relevant categories, our system seeks to enhance the utility and accessibility of cyber news, making it more actionable for cybersecurity professionals and organizations.

\subsection{Data}

We utilize Google News data via news alerts RSS feeds for our entire experiment. Initially, leveraging SME (Subject Matter Expertise), we configured the Google News Alerts with specific cyber terms such as 'data breach,' 'ransomware,' and 'cyberattack.' This setup generates a substantial number of news alerts daily, which we scrape and save to our database. We continually refine these feeds, adding new terms as they emerge in the cyber domain.

Over a period of one year and ten months, starting from early 2022 through to the end of October 2023, we have amassed a collection of approximately 265,000 cyber ralated articles. Each article is accompanied by metadata such as ID, link, publication date, and a headline, as detailed in Table~\ref{table:Data structure}. For our experiments, we focus exclusively on the headlines, which we have found to be sufficiently informative for our categorization challenge. Table~\ref{table:five_cyber_attack_categories} showcases various examples of news titles within each category.

\begin{table}[t]\centering
\caption{Data structure - Example of a single article entry from data theft feed}
\label{table:Data structure}
\begin{adjustbox}{width=0.48\textwidth}
\renewcommand{\arraystretch}{1.5}
\begin{tabular}
{|p{0.3in}|p{0.8in}|p{0.5in}|p{0.5in}|p{0.8in}|p{0.8in}|p{0.3in}|p{0.18in}|}
\hline
{\textbf{id}} &
{\textbf{link}} &
{\textbf{published\_ datetime}} &
{\textbf{updated\_ datetime}} &
{\textbf{headline}} &
{\textbf{content}} &
{\textbf{feed\_ name}} \\
\hline
UQPuJ TWfJ9 8oipe wHLcF Ef &
https://www.bnnb loomberg.ca/tesla-data-breach-blamed-on-insider-wrongdoing-impcted-75-000-1.1961383 &
2023-08-20 17:40:21. 000 &
2023-08-20 17:40:21. 000 &
Tesla Data Breach Blamed on 'Insider Wrongdoing' Impacted 75,000 - BNN Bloomberg &
(Bloomberg) -- Tesla Inc.'s May data breach impacted more than 75,000 people, included employee-related records and was a result of “insider ... &
data theft \\
\hline
\end{tabular}
\end{adjustbox}
\end{table}

\begin{table}[t]\centering
\caption{Five Cyber attack categories.}
\label{table:five_cyber_attack_categories}
\begin{adjustbox}{width=0.47\textwidth}
\renewcommand{\arraystretch}{1.2}
\begin{tabular}{|p{1.2cm}|l|}
\hline
{\textbf{Category}} & {\textbf{News Headlines}}\\
\hline
Recent & Royal Mail hit by cyber attack causing 'severe disruption' to services\\
Cyber & HHS compromised in massive MOVEit hack\\
attack & FTX says \$415 million of crypto was hacked\\
& ... \\ \hline
Future & Metro Bank Warns Against Rising Malware Attacks\\
Cyber & Apple issues 30-day warning to iPhone users\\
threat & Alarm raised over Mozilla VPN security flaw\\
& ... \\ \hline
& Cisco Releases A Fix For The Major ClamAV Antivirus Software Flaw\\
Cyber & Microsoft issues 75 patches, including three for zero-day\\
Prevention & Android 14 Will Block Malware With Enhanced Security Updates\\
& ... \\ \hline
& CommonSpirit Health sued over ransomware attack\\
Cyber & Meta hit with 390 mn euro fine over EU data breaches\\
Litigation & JPMorgan Must Face Lawsuit by Ray-Ban Maker over \$272 M\\
& ... \\ \hline
& This New McDonald's Hack Turns Sprite Into Cotton Candy Soda \\
Other & Samsung Galaxy Z Fold 5 can fix design flaw present in the brand's folding \\
& Our View: Google should have to answer for reckless site-blocking issues\\
& ... \\ \hline
\end{tabular}
\end{adjustbox}
\end{table}

\subsection{Theoretical Framework}

Given a news title comprising a sequence of \( N \) tokens \( w_1, w_2, \ldots, w_N \), we aim to determine a probability distribution \( \mathbf{P} \) over five cyber categories. This distribution is a function of the input tokens, represented as:

\begin{equation}
 \mathbf{P} = \mathbf{f}(w_1, w_2, \ldots, w_N)
\end{equation}

Where \( \mathbf{f} \) denotes the model that estimates the probability distribution  that satisfy the following condition:

\[
\begin{aligned}
&\sum_{i=1}^{5} P_i = 1, \\
&\text{where} \quad P_i \in \left\{ P_{\text{cyber\_attack}}, P_{\text{future\_threat}}, P_{\text{prevention}}, P_{\text{litigation}}, P_{\text{other}} \right\}
\end{aligned}
\]

The predicted final category is obtained through a decision function \( g \), which could be a simple maximization of the probabilities or a threshold-based approach:

\[
\text{Predicted Category} = g(\mathbf{P}), \quad \text{where} \quad g: \mathbf{P} \rightarrow \text{Category}
\]

This function evaluates the highest probability or applies a pre-defined threshold to determine the most likely category.

\section{Methodology}
\label{sec:methodology}

\subsection{Emerging Cyber Signal Discovery Module}

\begin{figure}
    \centering
    \includegraphics [width=6cm] {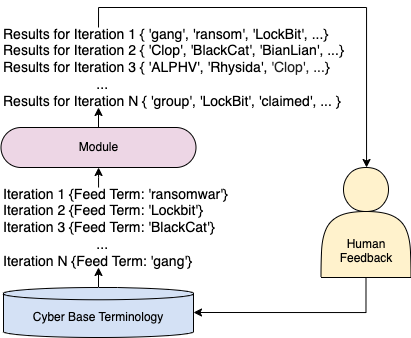}
    \caption{An illustration of Emerging Cyber Signal Discovery Module}
    \label{fig:Cyber_signal_detection}
\end{figure}

In response to increasing cyber attack frequencies, we've developed an Emerging Cyber Signal Discovery Module to identify and catalog new cyber attack terms for related news article retrieval. The module comprises three integral components: predefined cyber base terminologies, a tailored word vector model, and human feedback. Initially, a curated set of 30 cyber base terminologies is selected with the aid of subject matter expertise. Subsequently, each terminology is fed into the model, which, in turn, outputs related new terms meeting predefined criteria. Lastly, through a process of human feedback, the returned terms are meticulously reviewed and incorporated into the existing repository of cyber base terminologies. This systematic approach ensures the continual augmentation and refinement of the terminology lexicon to effectively adapt to the evolving landscape of cyber threats. An illustration of the module can see in Fig.~\ref{fig:Cyber_signal_detection}.

The word vector model that we use in this module is trained by all news articles that we collect at the moment using the skip-gram method \cite{mikolov2013distributed}. The training objective of the Skip-gram model is to find word representations that are useful for predicting the surrounding words in a sentence or a document. More formally, given a sequence of training words \( w_1, w_2, \dots, w_T \), the objective of the Skip-gram model is to maximize the average log probability

\begin{equation}
\frac{1}{T} \sum_{t=1}^{T} \sum_{-c \leq j \leq c, j \neq 0} \log p(w_{t+j} | w_t)
\end{equation}

where \( c \) is the size of the training context (which can be a function of the center word \( w_t \)). Larger \( c \) results in more training examples and thus can lead to a higher accuracy, at the expense of the training time.

The basic Skip-gram formulation defines \( p(w_{t+j} | w_t) \) using the softmax function:

\begin{equation}
p(w_O | w_I) = \frac{\exp({v'_{w_O}}^T v_{w_I})}{\sum_{w=1}^{W} \exp({v'_{w}}^T v_{w_I})}
\end{equation}

where \( v_w \) and \( v'_w \) are the "input" and "output" vector representations of \( w \), and \( W \) is the number of words in the vocabulary.

Utilizing the word vector model, we map words from the corpus into vectors within a hundred-dimensional vector space, formalizing the transformation as a function \( f: \text{word} \mapsto \mathbb{R}^{100} \). We then use cosine similarity to compute the similarity score between input cyber terms and other terms. The similarity score can be represented as:

\begin{equation}
\text{Similarity}(w_p, w_q) = \frac{v_{w_p} \cdot v_{w_q}}{\|v_{w_p}\| \|v_{w_q}\|}
\end{equation}

Here, $w_p$ and $w_q$ represent two different words, and $v_{w_p}$ and $v_{w_q}$ are their respective word vectors obtained from the skip-gram model. The dot product in the numerator measures the similarity in the orientation of the vectors, and the denominator normalizes the similarity by the magnitudes of the vectors. This cosine similarity metric provides a measure of the directional similarity between words in the vector space, ranging from 0 (completely dissimilar) to 1 (completely similar). This approach allows us to quantify the semantic relationships between cyber terms and other terms in our corpus, aiding in the analysis of the contextual associations between words.

\begin{figure}
    \centering
    \includegraphics [width=9cm] {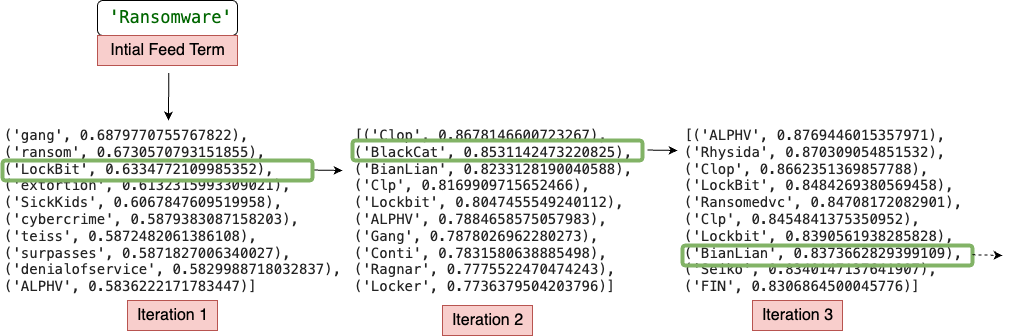}
    \caption{An illustration with Ransomware Feed Term }
    \label{fig:cyber_discovery}
\end{figure}

In our Cyber Signal Discovery module, we adhere to two principal criteria for term selection: (1) Candidate terms must exhibit at least a 60\% similarity score to the given cyber term. (2) Candidate terms must not duplicate or merely extend the existing entries in our cyber terminology database. Fig.~\ref{fig:cyber_discovery} illustrates this procedure using 'ransomware' as an input term. In the first iteration, terms such as 'LockBit' and 'SickKids' that score above the similarity threshold and are not already cataloged are identified. These are then subjected to human validation. For instance, 'LockBit' is approved and subsequently becomes a seed term for the next iteration, leading to the discovery of 'BlackCat' and 'BianLian' in subsequent rounds. This iterative process effectively enriches our database with relevant and emerging cyber terms.

\subsection{Random Forest Silver Labeling}

\vspace{2mm}

Silver labeling is an innovative technique situated between gold standard annotations and unsupervised predictions. It is particularly valuable in scenarios where acquiring labeled data is cost-prohibitive or logistically challenging. In our study, silver labels were generated to extend our training dataset, enabling us to train supervised models with a more substantial and diverse set of examples than would be feasible with manually annotated data alone.

We use the Random Forest algorithm for its ensemble ap-
proach that averages predictions from multiple decision trees, reducing overfitting risks. It is particularly suited for our small dataset size, outperforming gradient boosting methods like XGBoost\cite{Chen_2016} and LightGBM\cite{ke2017lightgbm} that typically require larger datasets to avoid learning noise as patterns. We implemented a Random Forest with 100 decision trees, optimized through cross-validation to ensure a balance between computational efficiency and predictive accuracy. The decision trees utilize Entropy as a measure to maximize split quality, defined as
\begin{equation}
E(S) = \sum_{i=1}^{c} -p_i \log_2 p_i
\end{equation}
where \( p_i \) represents the probability of an element belonging to class \( i \), and \( c \) is the number of classes.

Information Gain (IG), crucial for determining the most informative feature at each split, is used in our model to guide the decision-making process of the trees:
\begin{equation}
IG(D, A) = E(D) - \sum_{j=1}^{m} \frac{|D_j|}{|D|} E(D_j)
\end{equation}
Here, \( IG(D, A) \) is the information gain from partitioning dataset \( D \) with feature \( A \), \( E(D) \) is the dataset's entropy, \( m \) is the number of distinct values feature \( A \) can take, and \( E(D_j) \) is the entropy of the subset where \( A \) equals \( j \). The term \( \frac{|D_j|}{|D|} \) weights the entropy of each subset, indicating the proportion of instances in subset \( D_j \) post-split. Information gain quantifies the expected reduction in entropy upon learning the value of attribute \( A \).

\vspace{3mm}

\subsection{CANAL}

Central to our framework is CANAL (Cyber Activity News Alerting Language Model), a fine-tuned BERT \cite{devlin2019bert} model meticulously crafted for cyber news categorization. BERT's architecture includes multi-layer bidirectional transformer encoders, self-attention mechanisms, and feed-forward neural networks, which are adept at processing complex text.

The classifier layer is added atop the BERT model for classification tasks. The final hidden state from BERT, denoted by $h$, with weights $W$ and bias $b$, produces the output $y$ via:
\begin{equation}
y = f(W \cdot h + b)
\end{equation}
where $f$ is the softmax activation function. Fine-tuning with a lower learning rate using AdamW optimizer and a custom learning rate scheduler is crucial to retain the pre-trained knowledge.

The softmax function for the classifier output is:
\begin{equation}
\text{softmax}(y_i) = \frac{e^{y_i}}{\sum_{j} e^{y_j}}
\end{equation}
where $y_i$ is the output logit for class $i$.

\begin{figure}
    \centering
    \includegraphics [width=9cm] {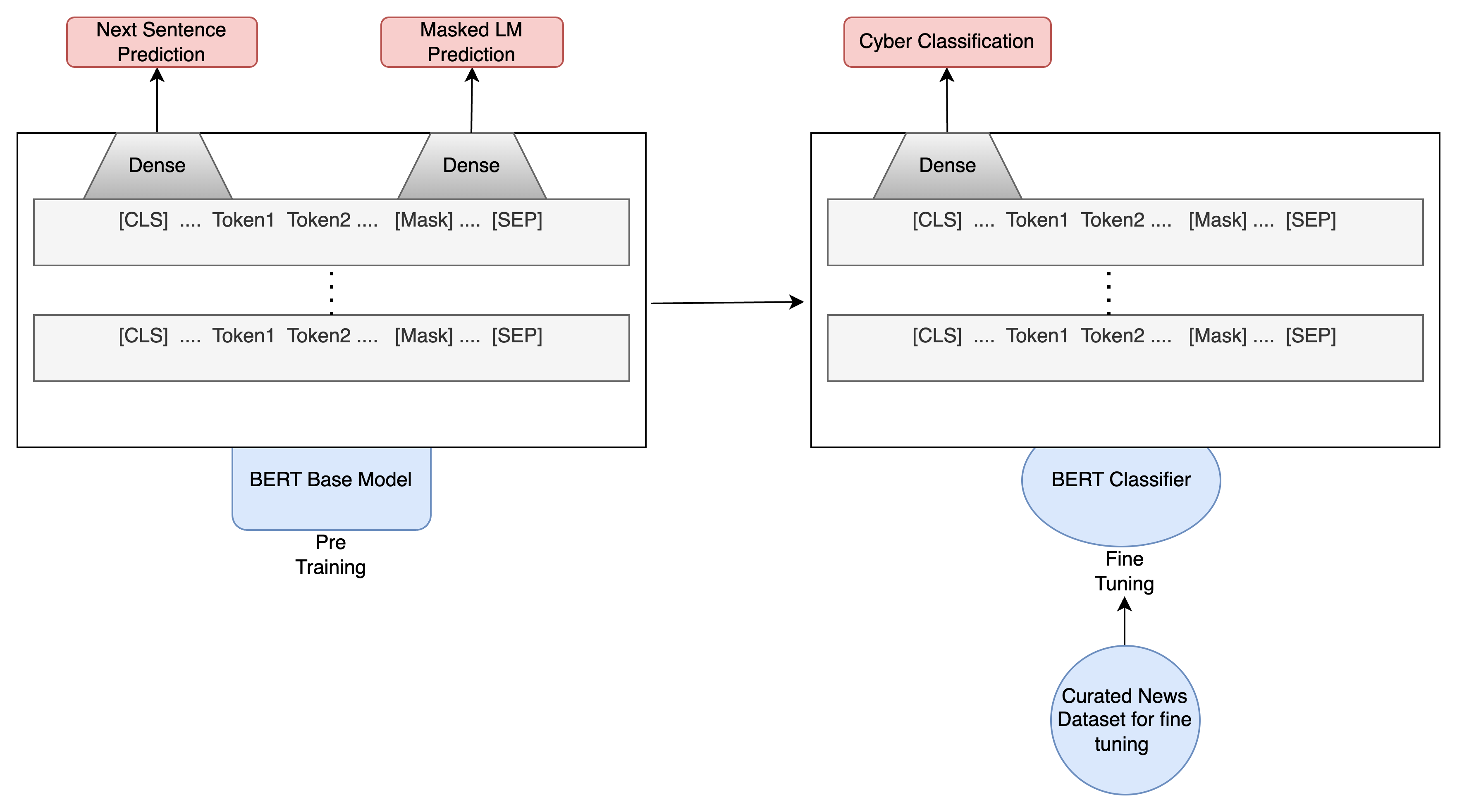}
    \caption{Illustration of CANAL with BERT Fine Tuning on cyber classification task.}
    \label{fig:BERT_finetuning architecture}
\end{figure}

The Fig.~\ref{fig:BERT_finetuning architecture} presents a schematic of CANAL. The left section outlines the BERT Base Model structure, showcasing its dual training objectives: Next Sentence Prediction and Masked Language Model (LM) Prediction. These components process sequences of tokenized inputs, each sequence initiated with a [CLS] token and interspersed with [SEP] tokens. In the right section, the BERT Classifier is depicted, which has been fine-tuned using a curated dataset specific to cyber-related news for enhanced classification accuracy. The classifier also processes tokenized inputs through dense layers, ultimately yielding categorized outputs. This schematic encapsulates the model's initial pre-training on diverse language data followed by its subsequent specialization through fine-tuning for cyber threat detection and classification.

We explore several other fine-tuning techniques to optimize CANAL's performance including Partial Finetuning with PEFT (Parameter-Efficient Fine-Tuning), and the integration of PEFT with LoRA (Low-Rank Adaptation).

\vspace{5mm} 

\subsubsection{Parameter-Efficient Fine-Tuning (PEFT)}

We explore Parameter-Efficient Fine-Tuning (PEFT) \cite{liu2022fewshot} for its efficiency in fine-tuning large LLMs. While full fine-tuning updates all parameters, partial fine-tuning in PEFT selectively freezes a portion of the model's weights while fine-tuning the rest.

The fine-tuning process for both full and partial parameter updates explores the performance impact on our multiclass classification task, providing insights into the trade-offs between computational efficiency and classification effectiveness.

\vspace{3mm}

\subsubsection{PEFT with LoRA}

\begin{figure}
\centerline{\includegraphics[width=0.25\textwidth]{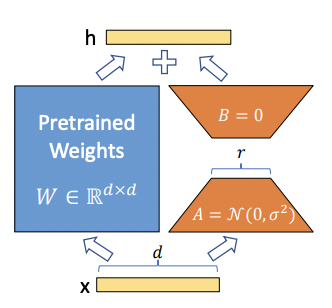}}
\caption{An illustration of LoRA from an original paper \cite{hu2021lora}}
\label{fig:lora}
\end{figure}

We also experiment with PEFT combined with Low-Rank Adaptation (LoRA) \cite{hu2021lora}. LoRA updates a pre-trained weight matrix $W_0$ with a low-rank decomposition $W_0 + \Delta W = W_0 + BA$, where $B \in \mathbb{R}^{d \times r}$ and $A \in \mathbb{R}^{r \times k}$, and the rank $r \ll \min(d, k)$. As shown in figure \ref{fig:lora}, during training, $W_0$ is frozen, while $A$ and $B$ contain trainable parameters. The modified forward pass with LoRA is:
\begin{equation}
h = W_0x + \Delta Wx = W_0x + BAx
\end{equation}

Our approach integrates BERT's architecture with PEFT and LoRA fine-tuning for effective cyber multiclassification, as demonstrated in our methodology.

\subsection{Generative Models}

In our investigation of cyber-related text classification, we benchmark our model against three state-of-the-art language models that are at the forefront of NLP advancements. At the time of writing, GPT-4 by OpenAI has emerged as a frontrunner, leading the field with its remarkable capabilities in generating human-like text and is accessible through the OpenAI API. Alongside GPT-4, Llama 13B and Zephyr 7B Beta have made significant strides in the open-source space, excelling across various NLP benchmarks. Below, we provide brief introductions to the technical details of each model.

\vspace{2mm}

\subsubsection{GPT-4}

GPT-4 \cite{openai2023gpt4}, an advanced multimodal model, stands out in the realm of natural language processing, particularly in processing image and text inputs to generate text outputs. Its proficiency is highlighted in complex scenarios, such as human-designed exams, where it notably scored in the top 10\% in a simulated bar exam. Surpassing earlier versions like GPT-3 and GPT-3.5, as well as other significant large language models (LLMs) including Cloude, Llama, and PaLM, GPT-4 demonstrates exceptional capabilities across multiple languages in standard NLP benchmarks. This versatility and robustness mark a new milestone in the evolution of language models.

\vspace{2mm}

\subsubsection{Llama 2 Chat (Unquantized)}

The \textit{Llama 2 Chat} \cite{touvron2023llama} model, a fine-tuned version of \textit{Llama 2}, is optimized for dialogue use cases and ranges from \(7 \times 10^9\) to \(7 \times 10^{10}\) parameters. Its training incorporates the pretraining of \textit{Llama 2} with public online sources, followed by supervised fine-tuning, and further refinement through Reinforcement Learning with Human Feedback (RLHF), using rejection sampling and Proximal Policy Optimization (PPO). The model adopts most architectural settings from \textit{Llama 1}, including the standard transformer architecture with modifications like pre-normalization using RMSNorm, the SwiGLU activation function \cite{shazeer2020glu}, and rotary positional embeddings (RoPE). Additionally, it features grouped-query attention and an increased context length, trained using the AdamW optimizer.

\vspace{2mm}

\subsubsection{Zephyr 7B Beta (Unquantized)}

The \textit{Zephyr} language model \cite{tunstall2023zephyr}, as detailed in the paper, represents a significant shift from the \textit{Llama 2 Chat} model in its approach to training and alignment with user intent. \textit{Zephyr}, specifically the \(7\)-billion parameter model \textit{Zephyr-7B}, is designed to be a smaller, more aligned model focusing on distilled supervised fine-tuning (dSFT) and distilled direct preference optimization (dDPO). Unlike \textit{Llama 2 Chat}, \textit{Zephyr} uses AI Feedback (AIF) from a set of teacher models for preference data, bypassing the need for human annotation and additional sampling. This approach enables rapid training in just a few hours, setting a new benchmark for \(7B\) parameter chat models and surpassing the performance of \textit{Llama 2 Chat} on certain benchmarks.

These three language models, GPT-4, Llama 13B (unquantized), and Zephyr 7B Beta (unquantized), serve as benchmarks in our evaluation, allowing us to assess the performance of our cyber-related text classification model in comparison to the industry-leading and open-source state-of-the-art models in the field.

\subsection{Entity Relevance Module}

We introduce the Entity Relevance Module as part of our system to enhance the processing of news articles by determining the contextual relevance of identified entities within the text. Unlike standard NER models that simply tag entities, this module assesses their contextual significance within the news titles.

For instance, in the statement "Cyber attacks rise, says Y Bank," a traditional NER model would identify "Y Bank" as an entity, but our module also evaluates its contextual relevance, recognizing that "Y Bank" is offering an opinion rather than being the central focus.

The model applies a sigmoid function to determine the probability of relevance:
\begin{equation}
P(\text{Class 1 - Relevant}) = \sigma(W \cdot \Phi(\text{input}) + b)
\end{equation}
where $\sigma$ denotes the sigmoid activation function, $W$ and $b$ are the model weights and bias, and $\Phi(\text{input})$ is the feature representation of the input.

While details of the training process are beyond this paper's scope, in summary, the Entity Relevance Module is trained on labeled data, producing probabilities that contribute to nuanced entity-centric analysis.

\vspace{5mm}

\section{Training and Evaluation Scheme }

\subsection{Train-Test Data}
\begin{itemize}
    \item \textbf{Train Set Composition:} The dataset utilized for training in this study was collated over a period extending from January 2022 to September 2023. This comprehensive dataset comprises a diverse array of samples a total volume of about 250000 samples.

    \item \textbf{Test Set Composition:}
    In assessing the performance of CANAL, exclusive data from October 2023 was employed for the evaluation. Specifically, a subset comprising 2000 articles from the articles of that month was sampled for testing.

\end{itemize}

\subsection{Data Labeling and Categorization}
\begin{itemize}
    \item \textbf{Gold Standard Dataset:} A subset of 600 samples from train set was meticulously labeled by domain experts to establish a 'Gold Standard' dataset. Special attention was paid to addressing class imbalance through stratified sampling and weighting classes inversely proportional to their frequencies in the input data. This dataset served as a benchmark for the initial Random Forest model training.

    \item \textbf{Silver Label Dataset using RF:}

     The Random Forest model, initially trained on the 600-sample 'Gold Standard' dataset, was then applied to the remaining data spanning from 2022 to September 2023. This step involved using the trained model to automate the labeling process, thereby generating 'Silver Labels' for a large volume of data. This approach enabled us with additional 8,000 records, which exhibited a high degree of certainty in their labeling.

    \item \textbf{BERT Model Fine-Tuning Data:} The selected sample set, enriched with both gold and silver labeled data, was employed for the fine-tuning of a BERT based model. This subset was specifically chosen to represent the population distribution accurately, ensuring that the model's training would be reflective of the diverse characteristics present within the larger dataset. This step was crucial for adapting the model to the specific nuances and characteristics of our dataset.

\end{itemize}

\subsection{Cyber Signal Discovery Module Training Scheme}

The Emerging Cyber Signal Discovery Module is executed on a monthly basis to identify novel cyber terms. Prior to each run, the word vector model undergoes updates through the latest news data.
We conclude the process either after 10 runs or when our stopping criteria are met. These criteria include: (1) words exhibiting a similarity score greater than 60\% to the specified cyber term, and (2) words that neither duplicate nor merely extend the entries present in our cyber terminology database. We halt the process at the occurrence of the first of these conditions. This periodic assessment ensures the continuous detection and incorporation of emerging cybersecurity terminology, contributing to the adaptability and efficacy of the system in capturing evolving cyber threats.

\subsection{Random Forest Training Scheme}

Our Random Forest (RF) model was trained on a Gold Standard dataset comprising 600 samples with expertly annotated labels to establish a robust initial model. The hyperparameters of the model are listed in Table~\ref{tab:hyperparameters}. To counter potential bias in silver label generation, we filtered our data retrieval with precise SQL queries, ensuring a representative dataset for training. This phase was critical for establishing a strong foundational model capable of further refinement and application on a larger dataset.

\begin{table}[t]
\centering
\caption{Random Forest Hyperparameters}
\label{tab:hyperparameters}
\begin{tabular}{@{}lc@{}}
\toprule
Hyperparameter & Value \\ \midrule
bootstrap & True \\
criterion & entropy \\
max\_depth & 10 \\
max\_features & auto \\
n\_estimators & 100 \\
random\_state & 42 \\
verbose & 0 \\
warm\_start & False \\ \bottomrule
\end{tabular}
\end{table}

\subsection{Training Scheme for BERT fine-tuning}

\subsubsection{Fine-Tuning Runs}
We conducted a series of four distinct BERT fine-tuning runs, each with a specific configuration aimed at exploring the effects of different levels of layer-wise training on model performance. The configurations for the fine-tuning runs were as follows:

\begin{itemize}
    \item \textbf{PeFT LoRA r=8}: This run involved a Parameter efficient fine tuning (PeFT) approach with a Low-Rank Adaptation (LoRA) with a rank of 8, which creates adapter weight matrices which are used in conjunction with complete model weights to work on classification task.
    \item \textbf{Last Layer + Classification Layer}: The second run was confined to training only the last layer of the BERT model in conjunction with the classification layer.
    \item \textbf{Last 2 Layers + Classification Layer}: The third run extended the training to include the last two layers of the BERT model alongside the classification layer.
    \item \textbf{Full BERT Fine-Tuning}: The final run encompassed a comprehensive fine-tuning of the entire BERT model.
\end{itemize}

\subsubsection{Gradient Freezing}
In our fine-tuning methodology, we employed gradient freezing for all layers except the ones undergoing fine-tuning. This technique not only requires less computational power but also reduces the time needed for model training compared to full model fine-tuning. Gradient freezing is a commonly adopted practice within the machine learning community for its efficiency in fine-tuning deep learning models.

\subsubsection{BERT Hyper parameters}
\begin{figure}[b]
\centering
\includegraphics[width=0.5\textwidth]{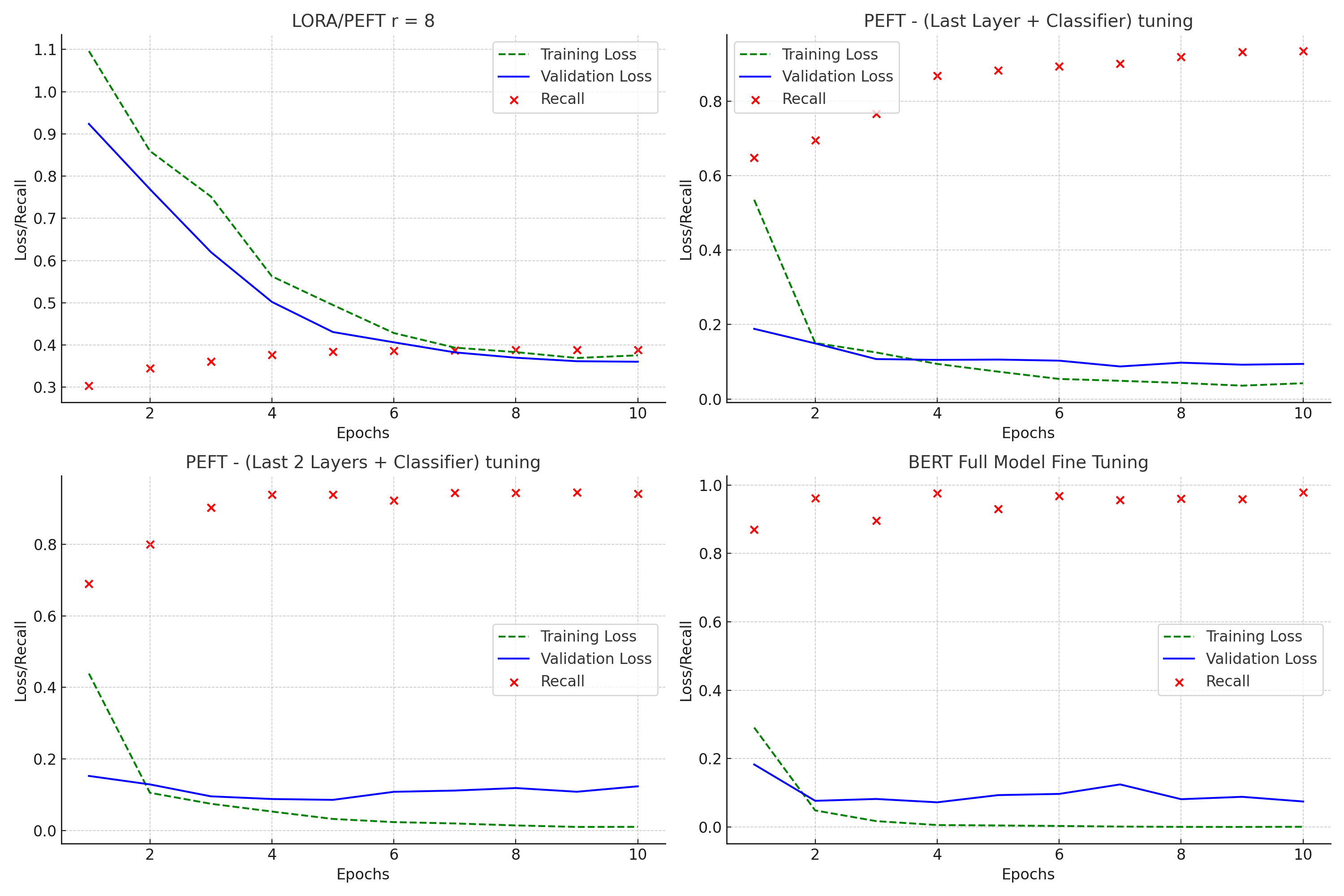}
\caption{Illustration of training and validation Cross-Entropy loss over 10 epochs}
\label{fig:recall_scores}
\end{figure}

The table \ref{tab:bert_hyperparameters} provides final hyper-parameters used for BERT model fine tuning. A comprehensive analysis of the training, validation cross entropy loss and recall trends across multiple epochs during the fine-tuning process of the BERT-based model is plotted in Fig.~\ref{fig:recall_scores}. The x-axis represents the number of training epochs, ranging from 1 to 10. The y-axis denotes the cross entropy loss values and recall values, which are computed for both the training and validation datasets.
\begin{table}[t]
\centering
\caption{BERT Training Hyperparameters}
\label{tab:bert_hyperparameters}
\begin{tabular}{lc}
\toprule
Parameter & Value \\
\midrule
learning\_rate & 2e-5 \\
per\_device\_train\_batch\_size & 8 \\
per\_device\_eval\_batch\_size & 8 \\
data\_seed & 727 \\
seed & 767 \\
save\_strategy & epoch \\
evaluation\_strategy & epoch \\
load\_best\_model\_at\_end & True \\
num\_train\_epochs & 10.0 \\
\bottomrule
\end{tabular}
\end{table}

\subsection{Prompt Engineering for LLMs}

In our approach to classifying cyber news into five categories, we iteratively refined prompt templates to enhance the understanding and performance of Language Large Models.

\begin{itemize}

\item  \textbf{Template 1, Zero Shot: Basic Instruction Set}

Our initial template provided only the basic instruction and the five categories, relying on the LLMs' inherent capabilities to interpret and classify the content correctly.

\begin{mdframed}[linewidth=0.8pt,linecolor=black,backgroundcolor=gray!8,roundcorner=8pt]
    \textbf{Template 1 Zero Shot:}

    \textit{You are a cyber analyst. You will be provided with a sentence of news regarding different entities. The entity could be an organization, location, place, person, or group. The sentence will be delimited with \#\#\#\# characters. Classify the sentence into one of five categories and output the category name only.}

    \textit{Five categories are:}
    \begin{enumerate}
        \item \textit{Recent Cyber Attack}
        \item \textit{Cyber Litigation}
        \item \textit{Cyber Attack Prevention}
        \item \textit{Future Cyber Threat}
        \item \textit{Other}
    \end{enumerate}
\end{mdframed}

\item  \textbf{Template 2, Zero Shot: Defined Criteria for Each Category}

To improve clarity, the second template incorporated specific criteria for each category, particularly emphasizing that the news should pertain to an identifiable entity in categories like Recent Cyber Attack, Litigation, Prevention, and Future Threat.

\begin{mdframed}[linewidth=0.8pt, linecolor=black, backgroundcolor=gray!8, roundcorner=8pt]
    \textbf{Template 2 Zero Shot:}

    \textit{You are a cyber analyst. You will ... Classify the sentence into one of the following five categories, now defined with specific criteria:}
    \begin{enumerate}
        \item \textit{\textbf{Recent Cyber Attack}: Sentences reporting on recent cyber attacks targeting entities.}
        \item \textit{\textbf{Cyber Litigation}: Sentences discussing legal actions, investigations, or charges related to cyber incidents.}
        \item \textit{\textbf{Cyber Attack Prevention}: Sentences highlighting positive actions, remedies, vulnerability fixes, and patches aimed at reducing the likelihood of future cyber risks.}
        \item \textit{\textbf{Future Cyber Threat}: Sentences addressing potential cyber risks and threats that organizations may face in the future.}
        \item \textit{\textbf{Other}: Sentences encompassing a diverse range of articles, including reports, studies, educational content, guidance materials, and miscellaneous topics related to cybersecurity.}
    \end{enumerate}
\end{mdframed}

\item  \textbf{Template 2, Few Shots: Inclusion of Examples}

To further assist the LLMs, we introduced examples in both one-shot and few-shot formats for each category. This approach was designed to further aid LLMs in grasping our categorization criteria, providing concrete instances as references.

\begin{mdframed}[linewidth=0.8pt, linecolor=black, backgroundcolor=gray!8, roundcorner=8pt]
    \textbf{Template 2 with Few Shots:}

    \textit{You are a cyber analyst. You will ... Classify the sentence into one of the five categories and output the category name only ...}

    \textit{Five categories are:}
    \begin{enumerate}
        \item \textit{\textbf{Recent Cyber Attack}: Sentences reporting on recent cyber attacks ...}
        \item ...
        \item \textit{\textbf{Other}: Sentences encompassing a diverse range of articles ...}
    \end{enumerate}

    \textbf{Examples for each category are:}
    \begin{lstlisting}[basicstyle=\ttfamily\footnotesize, breaklines=true]
    {"sentence": "Royal Mail hit by cyber attack causing 'severe disruption' to services.", "category": "Recent Cyber Attack"},
    ...
    {"sentence": "Meta hit with 390 mn euro fine over EU data breaches", "category": "Cyber Litigation"}
    \end{lstlisting}
\end{mdframed}

\end{itemize}

\section{Evaluation Results}

\subsection{Evaluation Metrics}

In this section, we outline the performance metrics utilized to assess the models in our multi-class classification task. CANAL, finetuned BERT model, is designed to optimize cross-entropy loss across five cyber categories in a multi-class categorization context. For a comparative analysis with other LLMs, where probability distributions are inaccessible, we adopt standard metrics such as Precision, Recall, F1-Score, and Accuracy. These metrics are chosen for their clarity in conveying performance insights. Here below is a quick introduction to all these metrics,

\textbf{Cross-Entropy Loss (Log Loss):}
Quantifies the difference between predicted and actual class probabilities.
\begin{equation}
\text{Log Loss} = -\frac{1}{N}\sum_{i=1}^{N}\sum_{j=1}^{M} y_{ij} \log(p_{ij})
\end{equation}
Here, $N$ is the number of instances, $M$ is the number of classes, $y_{ij}$ is 1 if instance $i$ is in class $j$, and $p_{ij}$ is the predicted probability.

\textbf{Precision:}
Evaluates the proportion of accurate positive predictions.
\begin{equation}
\text{Precision} = \frac{\text{True Positives}}{\text{True Positives + False Positives}}
\end{equation}

\textbf{Recall:}
Measures the model's ability to identify all positive instances.
\begin{equation}
\text{Recall} = \frac{\text{True Positives}}{\text{True Positives + False Negatives}}
\end{equation}

\textbf{F1-Score:}
Balances precision and recall, providing a comprehensive performance measure.
\begin{equation}
\text{F1-Score} = \frac{2 \cdot \text{Precision} \cdot \text{Recall}}{\text{Precision + Recall}}
\end{equation}

\textbf{Accuracy:}
Quantifies the overall correctness of the model's predictions.
\begin{equation}
\text{Accuracy} = \frac{\text{Number of Correct Predictions}}{\text{Total Number of Predictions}}
\end{equation}

\subsection{Evaluation Results}

\subsubsection{Evaluation of Cyber Signal Discovery Module}

The Emerging Cyber Signal Discovery Module operates with a human-in-the-loop moderation approach. During each iteration, the algorithm generates a set of signals on average, proportional to the quantity of novel cyber-related content in the news. Typically, 15-20\% of these signals are accepted through human validation. Compared to the exhaustive manual selection of cyber terminology, our module demonstrates remarkable efficiency in terms of time savings and in discerning emerging cyber attack-related terms.

\begin{figure}
\centering
\includegraphics[width=0.4\textwidth]{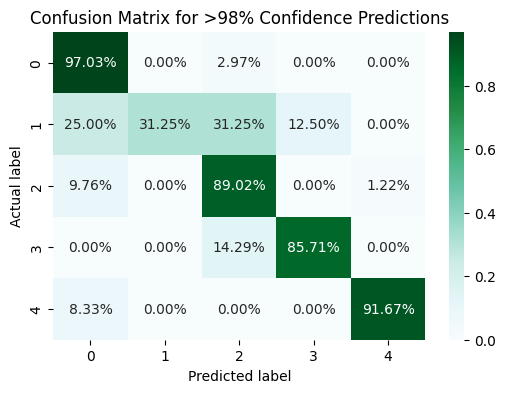}
\caption{Confusion matrix for a Random Forest classifier, with instances exceeding a 0.98 probability threshold. Categories are represented with {0: ’Other’, 1: ’Prevention’, 2: ’Recent Cyber Attack’, 3: ’Future Threat’, 4: ’Litigation’}}
\label{fig:rf_98_880_sample}
\end{figure}

\begin{figure}[b]
\centering
\includegraphics[width=0.5\textwidth]{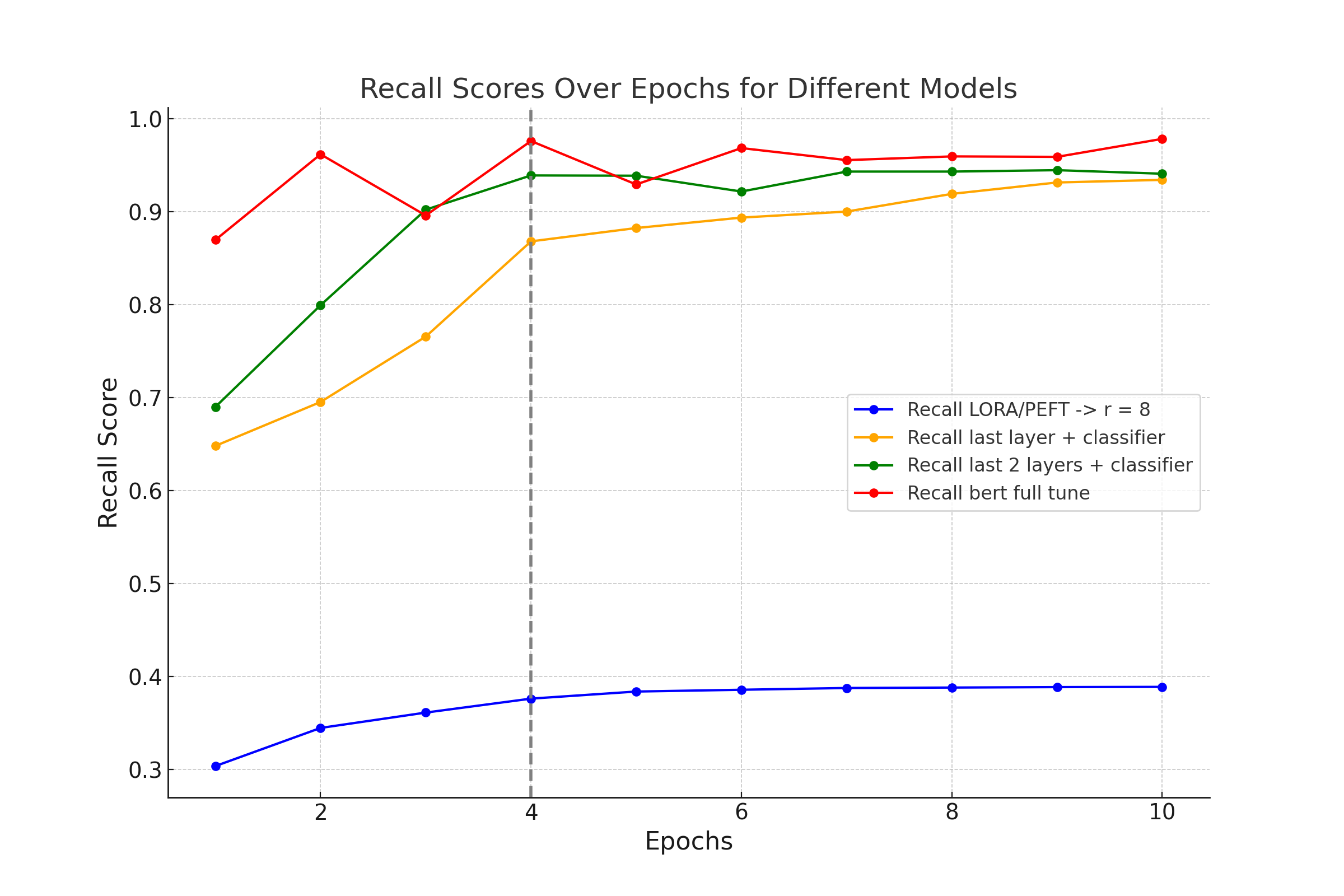}
\caption{Recall scores for BERT Fine-Tuning Schemes}
\label{fig:recall_scores_}
\end{figure}

\subsubsection{Evaluation of Random Forest Silver Labeling}

We applied various probability thresholds to the Random Forest classifier, ultimately settling on a 0.98 cutoff for sample inclusion. This threshold, rigorously chosen, ensures that only predictions with a confidence level of 0.98 or higher are considered. We ensured that this approach mirrors the distribution observed in the full dataset, allowing us to cultivate a sample that closely aligns with gold standard data quality. The resulting subset, approximately 8,000 records strong, showcases a high level of labeling confidence. Figure~\ref{fig:rf_98_880_sample} presents a confusion matrix for the Random Forest classifier, demonstrating its accuracy when operating above this stringent confidence threshold.

\subsubsection{Evaluation of BERT Fine-Tuning schemes}

After examining the loss in Fig.~\ref{fig:recall_scores} and  the recall in Fig.~\ref{fig:recall_scores_} at checkpoint 4 (epoch 4), the full fine-tuning of the BERT model demonstrated the best performance compared to other methods. While the focus was on absolute performance in which full fine-tuning excelled, it is worth noting that other configurations, such as fine-tuning only the \textbf{last layer plus the classifier} and the \textbf{last two layers plus the classifier}, also showed promising results. These strategies may be preferred in scenarios where computational resources are limited or when working with very large models, where full fine-tuning would be too resource-intensive.

\subsubsection{Evaluation of Expensive LLMs}

\begin{figure}
    \centering
    \includegraphics [width=8cm] {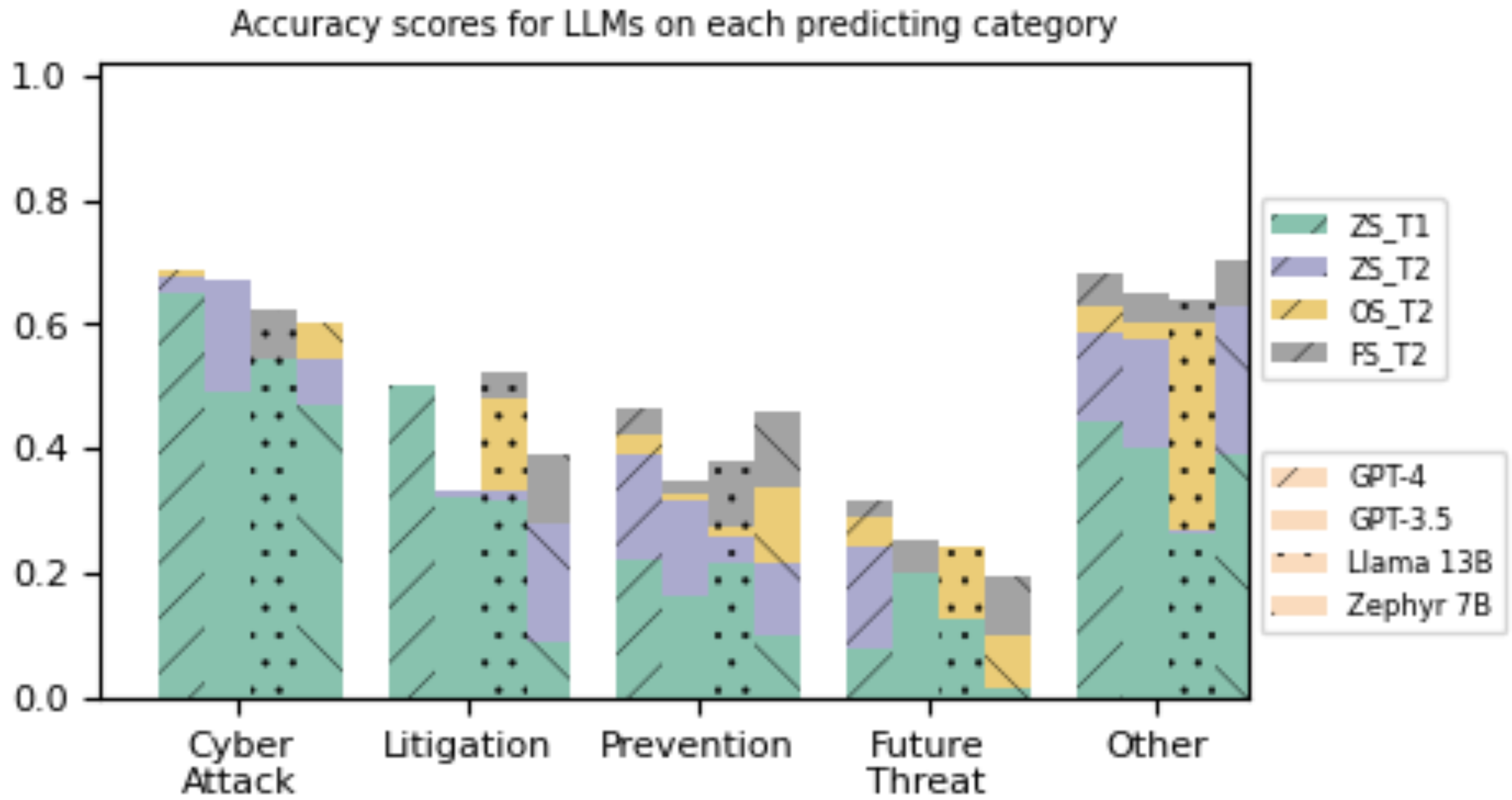}
    \caption{Accuracy Improvements with Prompt Engineering}
    \label{fig:plot_group_stacked_bar_chart_ac}
\end{figure}

\begin{figure}[b]
    \centering
    \includegraphics [width=8cm] {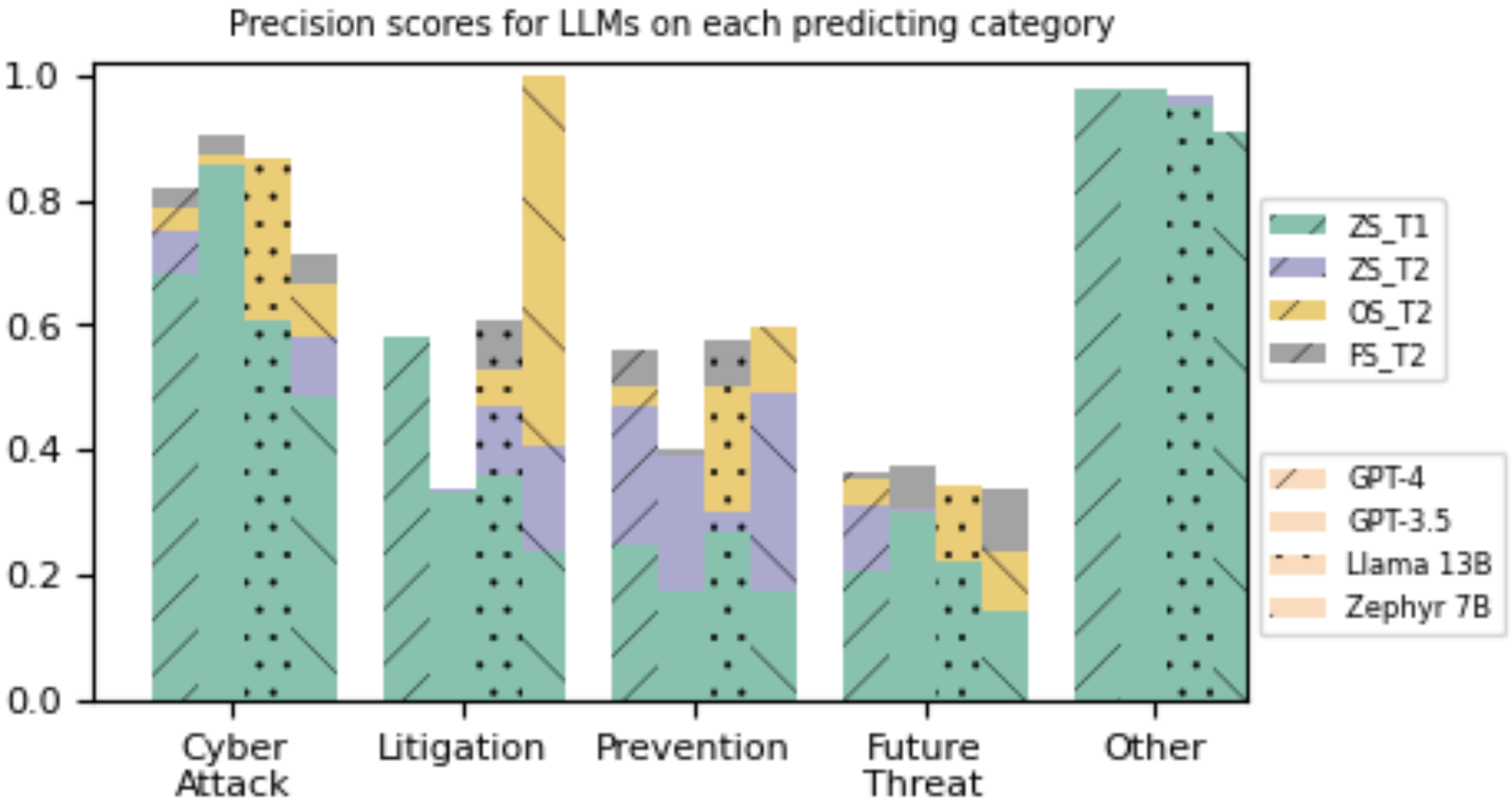}
    \caption{Precision Improvements with Prompt Engineering}
    \label{fig:plot_group_stacked_bar_chart_pc}
\end{figure}

\begin{figure}
    \centering
    \includegraphics [width=8cm] {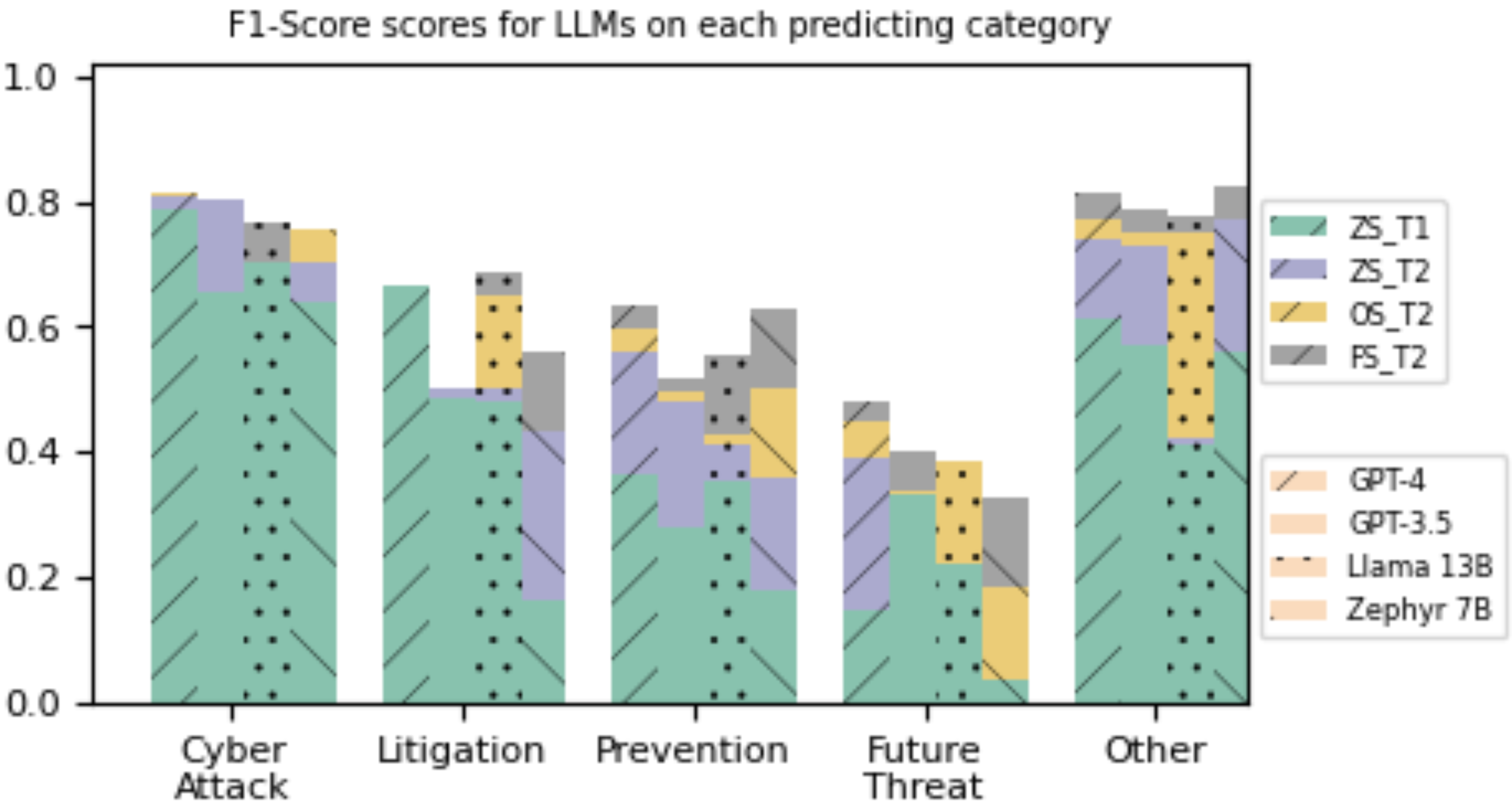}
    \caption{F1 scores Improvements with Prompt Engineering}
    \label{fig:plot_group_stacked_bar_chart_f1}
\end{figure}

\begin{figure}[b]
    \centering
    \includegraphics [width=8cm] {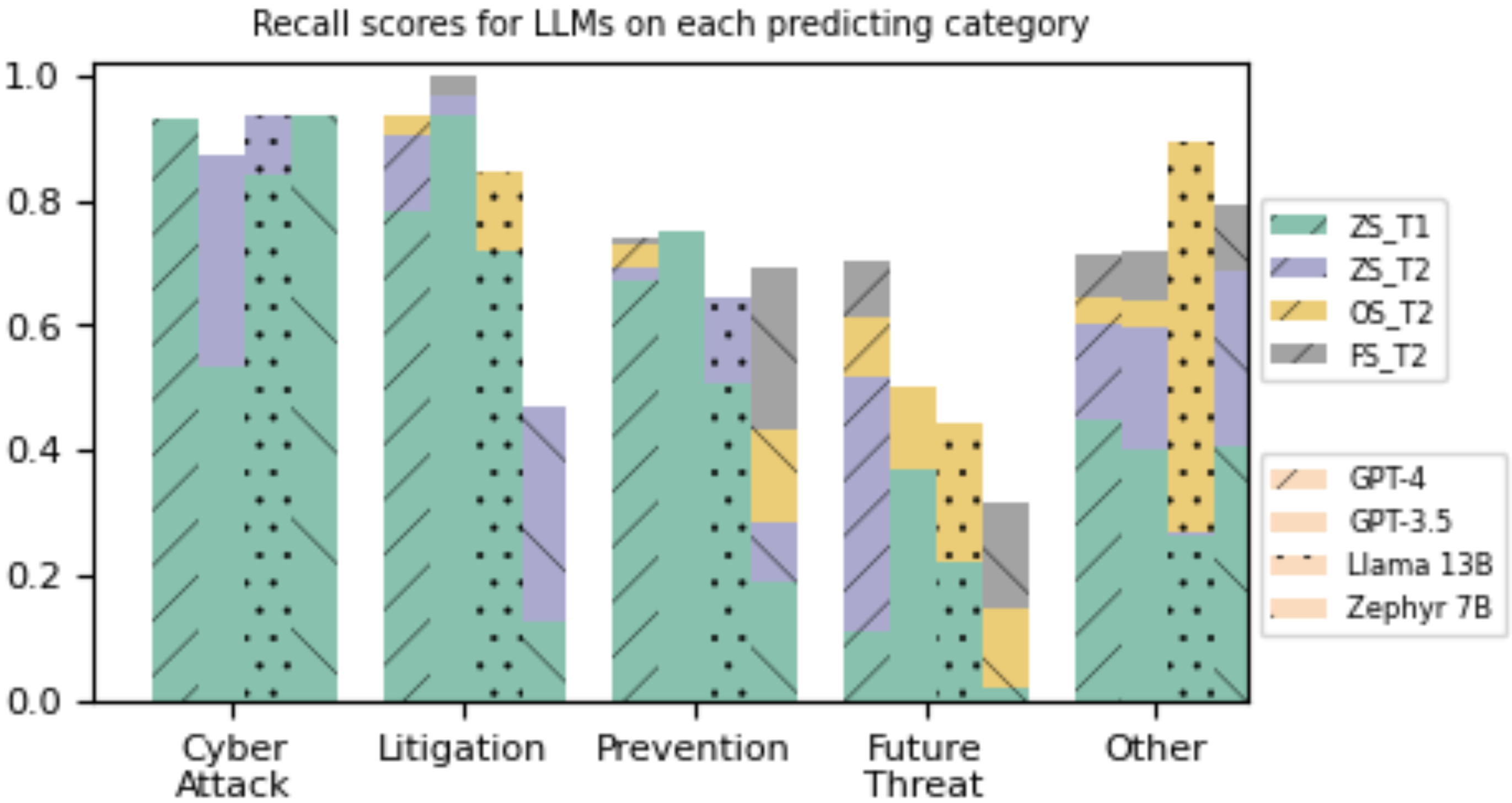}
    \caption{Recall Improvements with Prompt Engineering}
    \label{fig:plot_group_stacked_bar_chart_rc}
\end{figure}

Fig.~\ref{fig:plot_group_stacked_bar_chart_ac} to Fig.~\ref{fig:plot_group_stacked_bar_chart_rc} illustrate the performance of sophisticated language models (LLMs) in predicting various categories using different matrices. The abbreviations in these figures are as follows: ZS for Zero Shot, OS for One Shot, FS for Few Shot, T1 for Template 1, and T2 for Template 2. Upon examination, both GPT-4 and GPT-3.5 demonstrate a robust ability to identify cyber attacks, supported by high precision and recall. However, they encounter challenges in litigation, prevention, and future threat scenarios, where recall is exceptionally high but precision is relatively low. This suggests a tendency to over-identify instances in litigation, prevention, and future threat, potentially leading to more false positives.

In Fig.~\ref{fig:plot_group_stacked_bar_chart_pc} and Fig.~\ref{fig:plot_group_stacked_bar_chart_rc}, Llama 2 exhibits strong recall but lower precision in litigation and prevention scenarios under the template 1 Zero Shot, mirroring the behavior observed in GPT-4 and GPT-3.5. Notably, when presented with additional examples, there is a noticeable improvement in precision, signifying the model's ability to learn from these examples and reduce false positive predictions in both litigation and prevention categories.

In the context of Zephyr 7B, it is noteworthy that the application of prompt engineering yields more benefits compared to other LLMs. This insight is illustrated in Fig.~\ref{fig:plot_group_stacked_bar_chart_f1}, where when prompt engineering is applied, its F1 scores for litigation, prevention, future threat, and other categories show more significant improvement than those of other LLMs. Additionally, it is noticed that the results produced by Zephyr deviate from the desired structure, requiring extra time and effort for the extraction of predictions.

In our comparative analysis, GPT-4 emerged as a front-runner among the LLMs, yet the overall performance of these models did not meet our expectations. As Table~\ref{tab:model_performance} shows, none achieved an F1 score above 82\% in any category. This suggests that despite the use of prompt engineering to enhance task understanding, these models still face challenges in fully grasping the tasks. Notably, a closer examination of false positives revealed a tendency for incorrect predictions in samples with ambiguous content or lacking specific entity references, even with explicit instructions. This finding points to a critical area for improvement in terms of the LLMs ability to discern and interpret nuanced or incomplete information.

\subsubsection{CANAL vs Other Expensive LLMs}

\begin{table}[t]
\centering
\caption{Model Performance Comparison}
\label{tab:model_performance}
\small 
\setlength{\tabcolsep}{2pt} 
\begin{tabular}{lcccccc}
\toprule
\textbf{Model} & \textbf{Category} & \textbf{Accuracy} & \textbf{Precision} & \textbf{Recall} & \textbf{F1-Score} \\
\midrule
GPT-3.5 & Cyber Attack & 58.67 & \textbf{90.26} & 62.63 & 73.95 \\
FS & Litigation    & 27.35 & 27.35 & \textbf{100.00} & 42.95 \\
   & Prevention    & 35.00 & 39.87 & \underline{74.12} & 51.85 \\
   & Future Threat & 25.00 & \underline{37.70} & 42.59 & 40.00 \\
   & Other         & 65.25 & 87.93 & 71.66 & 78.97 \\
\midrule
GPT-4  & Cyber Attack & \underline{66.67} & 81.78 & \underline{78.29} & \underline{80.00} \\
FS       & Litigation    & 42.86 & 44.12 & \underline{93.76} & 60.00 \\
         & Prevention    & \underline{46.67} & 55.75 & \textbf{74.13} & \underline{63.64} \\
         & Future Threat & \underline{31.40} & 36.19 & \textbf{70.37} & \underline{47.80} \\
         & Other         & 68.39 & \textbf{94.14} & 71.43 & 81.23 \\
\midrule
Llama 13B & Cyber Attack & 62.43 & \underline{87.93} & 71.66 & 78.97 \\
FS        & Litigation    & \underline{52.08} & 60.98 & 78.13 & \underline{68.49} \\
          & Prevention    & 38.14 & \underline{57.69} & 52.94 & 55.21 \\
          & Future Threat & 18.64 & 25.58 & 40.74 & 31.43 \\
          & Other         & 63.75 & 87.68 & 70.02 & 77.86 \\
\midrule
Zephyr 7B    & Cyber Attack & 58.81 & 71.15 & 77.22 & 74.06 \\
FS           & Litigation    & 38.89 & \underline{77.78}
             & 43.75 & 56.00 \\
             & Prevention    & 45.74 & 57.28 & 69.41 & 62.77 \\
             & Future Threat & 19.54 & 34.00 & 31.48 & 32.69 \\
             & Other         & \underline{70.42} & 86.45 & \underline{79.16} & \underline{82.64} \\
\midrule
CANAL & Cyber Attack & \textbf{81.44} & 83.69 & \textbf{96.80} & \textbf{89.77} \\
     & Litigation    & \textbf{88.24} & \textbf{93.75} & 93.75 & \textbf{93.75} \\
     & Prevention    & \textbf{60.82} & \textbf{83.10} & 69.41 & \textbf{75.64} \\
     & Future Threat & \textbf{47.37} & \textbf{90.00} &  \underline{50.00} & \textbf{64.29} \\
     & Other         & \textbf{86.37} & \underline{93.35} & \textbf{92.04} & \textbf{92.69} \\
\bottomrule
\vspace{-6pt}
\end{tabular}
{\raggedright Note. The best results in each category are highlighted in bold, while the second-best results are underlined. FS stands for Few Shots.\par}
\end{table}

In the comparative evaluation between our framework, CANAL, and other expensive LLMs, CANAL emerges as the superior performer across all five assessed categories. For categories with very few samples, such as prevention and the future threat category, we expected these highly capable LLMs to perform well due to their training on large datasets. However, CANAL outperforms the expensive LLMs in these categories. In Table~\ref{tab:model_performance}, the accuracy of CANAL for the prevention category is approximately 13\% higher than that of GPT-4 Few Shots, and the accuracy of CANAL for the future threat category is around 16\% higher than that of GPT-4 FS.

Conversely, in categories with a more substantial number of training samples, such as recent cyber attack and other category, CANAL exhibits even more exceptional performance than other LLMs. In the recent cyber attack category, where GPT-4 attains the highest accuracy among all high parameter LLMs with approximately 67\%, CANAL surpasses this performance with an impressive accuracy of about 81\%. Similarly, in the Other category, where Zephyr 7B leads with around 70\% accuracy among LLMs, CANAL outshines with a superior accuracy of about 86\%. These findings underscore CANAL's robust and consistently superior performance across diverse categories, reinforcing its efficacy in handling nuanced language tasks.

In Table~\ref{tab:running_time_comparison}, we provide an empirical comparison of inference time and cost for processing 10,000 articles between CANAL and other LLMs. Our results show that CANAL is highly efficient and cost-effective. Cost estimations are based on the current OpenAI API pricing \cite{openai_pricing} and the lowest available A100 \& T4 usage rates, approximately \$1/hour \&
\$0.071/hour respectively\cite{Databricks_pricing} \cite{A100_price} at the time of writing. CANAL not only processes data faster but also maintains minimal operational costs, emphasizing its viability for large-scale applications.

\begin{table}[ht]
\centering
\caption{Comparison of Models for Processing 10,000 Articles }
\label{tab:running_time_comparison}
\small 
\setlength{\tabcolsep}{1.5pt} 
\begin{tabular}{@{}lS[table-format=1.2]lS[table-format=2.1]@{}}
\toprule
\textbf{Model} & {\textbf{Inference Time (hr)}} & \textbf{Infrastructure} & {\textbf{Inference Cost (\$)}} \\
\midrule
GPT-3.5 & 1.33 & OpenAI API & 4.4 \\
GPT-4 & 2.50 & OpenAI API & 84.5 \\
Llama 13B & 7.83 & A100 (48 GB) & 10.2 \\
Zyphyr 7B & 2.83 & A100 (48 GB) & 3.0 \\
CANAL & 0.067 & T4 (16 GB) & 0.0047 \\
\bottomrule
\end{tabular}
\end{table}

\vspace{5mm}
\subsubsection{CANAL in action}

\begin{table}[t]
\centering
\caption{Example classification snippets.}
\label{tab:text_classification}
\begin{tabular}{p{4cm} p{4cm}}
\toprule
\textbf{Cyber News} &
\textbf{CANAL + Cyber Signal Detection + Entity Relevance} \\
\midrule
McLaren Health Care Facing 3 Lawsuits in Ransomware... & \textcolor{red}{McLaren Health Care} Facing 3 Lawsuits in \textcolor{blue}{Ransomware Hack} - \textcolor{darkgreen}{\textit{Class - Litigation}} \\
Seiko confirms thousands of user accounts were ... & \textcolor{red}{Seiko} confirms thousands of user accounts were breached in \textcolor{blue}{cyberattack}  - \textcolor{darkgreen}{\textit{Class - Cyber Attack}} \\
Microsoft fixes over 100 vulnerabilities, 2 act... & \textcolor{red}{Microsoft} fixes over 100 \textcolor{blue}{vulnerabilities}, 2 actively exploited \textcolor{blue}{bugs}  - \textcolor{darkgreen}{\textit{Class - Prevention}} \\
Buffer overflow bug gives root on potentially millions... & \textcolor{blue}{Buffer overflow bug} gives root on potentially millions of \textcolor{red}{Linux} boxes - The Stack - \textcolor{darkgreen}{\textit{Class - Future Threat}} \\
Mastercard introduces new grocery delivery and ... & \textcolor{red}{Mastercard} introduces new grocery delivery and streaming perks - CNBC - \textcolor{darkgreen}{\textit{Class - Other}} \\
\bottomrule
\end{tabular}
\end{table}

In Table \ref{tab:text_classification}, we exhibit the culmination of our multifaceted analysis pipeline. Our dual approach begins with key element extraction: identifying relevant entities (marked in red) via an entity relevance model and pinpointing cyber signals (highlighted in blue) through our continually updated cyber taxonomy. Subsequently, CANAL is deployed for cyber news classification. This integrated process enables us to construct in-depth cyber risk profiles, leveraging news data tailored to specific entities.

\vspace{5mm}
\section{Conclusion}

We have successfully demonstrated that our framework, uniquely trained using a silver labeling approach with just 600 manually labeled data points, excels in 5-class cyber news categorization. CANAL not only outperforms current top Large Language Models in this specific task but also provides a more cost-efficient and empirical approach. Additionally, our detailed comparison reveals how existing industry LLMs perform in cyber categorization tasks and suggests that their efficacy can be further enhanced through prompt engineering techniques. Furthermore, we showcased the effectiveness of our Cyber Signal Discovery module in identifying emerging cyber signals. Future research may focus on leveraging advancements in cost-effective and smaller-sized empirical LLMs, specifically tailored to the dynamic requirements of cyber categorization.




\vspace{12pt}


\end{document}